\documentstyle[nato,epsf,numreferences,graphicx,amssymb,cite]{crckapb} 

\newcommand{\Z}{{\mathbb{Z}}}

\newcommand{\eq}{\begin{equation}}
\newcommand{\en}{\end{equation}}
\newcommand{\um}{\frac12}
\newcommand{\bra}{\langle}
\newcommand{\ket}{\rangle}

\begin{opening}
\title{CENTER VORTICES, MAGNETIC CONDENSATE\protect\\
 AND CONFINEMENT IN A SIMPLE GAUGE SYSTEM}
\author{F. GLIOZZI }
\author{M. PANERO}  

\institute{ Dipartimento di Fisica Teorica dell'Universit\`a di
  Torino,\\
Istituto Nazionale di Fisica Nucleare, Sezione di Torino,\\
  via P. Giuria 1, I 10125 Torino, Italy}
\author{P. PROVERO}
\institute{ Dipartimento di Scienze e Tecnologie Avanzate\\
Universit\`a del Piemonte Orientale, Alessandria, \\
Istituto Nazionale di Fisica Nucleare,\\ gruppo collegato di Alessandria,
  I 15100 Alessandria, Italy}

\end{opening}

\runningtitle{CENTER VORTICES AND CONFINEMENT}

\begin{document}

% The \begin{document} command comes after the \end{opening}
% command.
 
\begin{abstract}
 The confining mechanisms of 't Hooft and Mandelstam
have a simple microscopic realization in 3D $\Z_2$ gauge theory: the
center vortex and the magnetic monopole condensation are associated,
in the set of configurations contributing to the confining phase, to
the presence of two kinds of infinite clusters. These generate the area
law of the large Wilson loops and  the universal finite size effects
produced by the quantum fluctuations of the bosonic string describing
the infrared behavior of the flux tube. 
\end{abstract}

\section{Introduction}
Two kinds of topological excitations were proposed by 't Hooft as 
possible composite fields condensing in the disordered phase of  
gauge systems  to produce confinement: magnetic monopoles
\cite{'tHooft:1981ht} and center vortices\cite{tHooft:1977hy}.

Condensation of magnetic monopoles implies dual superconductivity of
the vacuum and gives the well-known physical picture of confinement 
in terms of dual Abrikosov vortices which describe the flux tubes.

Center vortices are string-like structures which are created
by gauge transformations with a non-trivial homotopy. Their
condensation produces a very efficient disordering of the gauge
configurations.
 
In the last few years, many lattice studies have been made on these
two subjects in the last
few years (see for instance the two review talks of  Di Giacomo 
\cite{Giacomo:2002mm} and Olejn\'\i k  \cite{Faber:2002ib} at this 
Conference).

These two kinds of confining mechanisms have a particularly simple
implementation in the 3D  $\Z_2$  gauge model, where it is possible
to study this phenomenon not only in terms of the effective degrees of
freedom, using for instance the underlying string description of the 
infrared behavior of the flux tube, but also at a microscopic level,
i.e. in terms of $\Z_2$ lattice configurations: it is possible to pick
out the  degrees of freedom which are relevant for the infrared
dynamics of the confining phase. 

It turns out that the condensation of
the above-mentioned topological excitations manifests itself at the
microscopic level with the appearance of two (complementary) infinite
clusters of links in the dual lattice. Both produce not only area law decay
of the vacuum expectation value of large Wilson loops, but also the universal
finite size effects which are predicted by the free bosonic string
picture of the flux tube.

\section{Monopole condensation}
The three-dimensional $\Z_2$ gauge model on a cubic lattice is defined
by the action
\begin{equation} 
S(\beta)=-\beta\sum_{\Box}U_{\Box}~~~~,~ U_{\Box}=\prod_{\ell\,\in\Box}U_\ell
\end{equation} 
where $U_\ell=\pm1$ is the $\Z_2$ gauge field associated to the links
$\ell$ of the lattice $\Lambda$. 
The magnetic monopoles live in the dual lattice $\tilde{\Lambda}$. In
order to create a monopole in a site $\tilde{x}\in\tilde{\Lambda}$
corresponding to the center of an elementary cube of $\Lambda$ it is
sufficient to draw an arbitrary, continuous line $\gamma(\tilde x)$ 
joining $\tilde{x}$ to $\infty$ and flipping the sign of the coupling 
of the plaquettes crossed by $\gamma$. This flipping is generated by the
non-local operator
\begin{equation} 
\Psi_\gamma(\tilde x)=\exp(-2\beta\sum_{\Box\in\gamma}U_\Box)~~.
\end{equation} 
As a consequence, the flux across any closed
surface around $\tilde{x}$ is equal to -1: this monopole field
$\Psi_\gamma(\tilde x)$ creates
one unity of  $\Z_2$ flux.
Thus the monopole is associated to a point-like topological defect in 
the ensemble of the gauge configurations.

Monopole condensation occurs when $\Psi$ acquires a vacuum expectation 
value different from zero:
$$ {\rm monopole~ condensation} \Longleftrightarrow\, \bra\Psi\ket\not=0 ~ .$$
Why this condensation should imply confinement?
A useful piece of information comes from the Kramers- Wannier duality.
This transformation maps the $\Z_2$ gauge theory into the Ising model
 described by the Hamiltonian
\eq
H(\beta)=-\tilde\beta\sum_{\bra\tilde x\tilde y\ket}\sigma_{\tilde x}
\sigma_{\tilde y}~~~,
~\tilde {\beta}=-\um\log\tanh\beta
\en
with $\bra\tilde x\tilde y\ket$ ranging over the links of 
$\tilde{\Lambda}$ and the spin variable 
$\sigma_{\tilde x}$ taking the values $\pm 1$.  As a consequence one 
can easily show that 
\eq
\bra\Psi_\gamma(\tilde x)\ket=\bra\sigma_{\tilde x}\ket_{Ising}
\en
Thus the magnetic condensation is associated with the spontaneous $\Z_2$ 
symmetry breaking of the Ising model.

The same transformation maps the Wilson loop $W(C)$ associated
 to any closed curve $C\in\Lambda$ into the corresponding 't Hooft loop 
$\widetilde{W}(C)$
of the Ising model:
\eq
\bra W(C)\ket= \bra \widetilde{W}(C)\ket_{Ising}
\en
with
\eq
\widetilde{W}(C)=\exp\left(-2\tilde{\beta}\sum_{\bra ij\ket\in\Sigma}
\sigma_i\sigma_j\right)~~,
~\; \partial\Sigma=C
\en
where $\Sigma$ is an arbitrary surface bounded by $C$.

$\widetilde{W}(C)$ creates an elementary $\Z_2$ flux along $C$ which 
manifests itself as a topological defect or a frustration in the
ensemble of the Ising configurations:
the product of the link variables along any loop $\widetilde{C}\in\widetilde
{\Lambda}$ having 
an odd linking number with $C$ is equal -1.

Confinement requires that $\bra\widetilde{W}(C)\ket_{Ising}$  should decay
with an area law in the ordered phase of the Ising model.
To understand this property at a microscopic level it is convenient to 
resort to
the  Fortuin Kasteleyn (FK) random cluster representation of the model:
\eq
Z=\sum_{\{\sigma\}}e^{-H(\beta)}=\sum_{G\subseteq\Lambda}v^{b_G}2^{c_G}~~~,
\label{Z}
\en
where $v=e^{\tilde\beta}-1$, the summation is over all 
spanning 
subgraphs $G\subseteq\Lambda$. $b_G$ is the number of links of $G$, 
called active bonds,
and $c_G$ is the number of connected components, called FK clusters. 
The ordered phase
 is characterised by the presence of an infinite, percolating  FK 
cluster.

When there are frustrations in the system (for instance those generated 
by $\widetilde{W}(C)$)
 the summation in Eq.\ref{Z} is constrained. Only those FK clusters are 
allowed for which
no closed path within the cluster is linked to an elementary $\Z_2$ 
flux \cite{Caselle:1999hy}. This is a 
microscopic realization of a sort of dual Meissner effect: the FK 
clusters behave like
pieces of dual superconducting matter of type I and no $\Z_2$ flux can 
go through them.
Introducing a projector $\varpi_C(G)$ on the configurations 
$G\subseteq\Lambda$ which takes 
the value 0 whenever there is  a FK cluster (at least)  linked to 
$C$, and the value 1  in all the other cases, yields the very useful identity
\eq
\bra W(C)\ket=\bra\varpi_C\ket_{Ising}=\frac{\rm number~ of~ allowed~ 
config.}{\rm total~number~ of~ config.}~~.
\en 
Assume that there are only FK clusters of finite size ( this is the 
case of the symmetric phase). If $C$ is much larger than the mean 
size of the clusters, 
the configurations contributing to $W(C)$, i.e. those with $\varpi_C(G)
=1$, are
 characterised by the fact that there is no cluster linked to $C$. The 
weight of this class
of configurations, when compared with the total ensemble 
$G\subseteq\Lambda$ is 
clearly suppressed by a factor $e^{-\alpha\, p(C)}$, where $p(C)$ is the 
length of $C$. 
\begin{figure}
%\begin{center}
%\input{largest.tex}
%\end{center}
\centering
\includegraphics[width=0.8\textwidth]{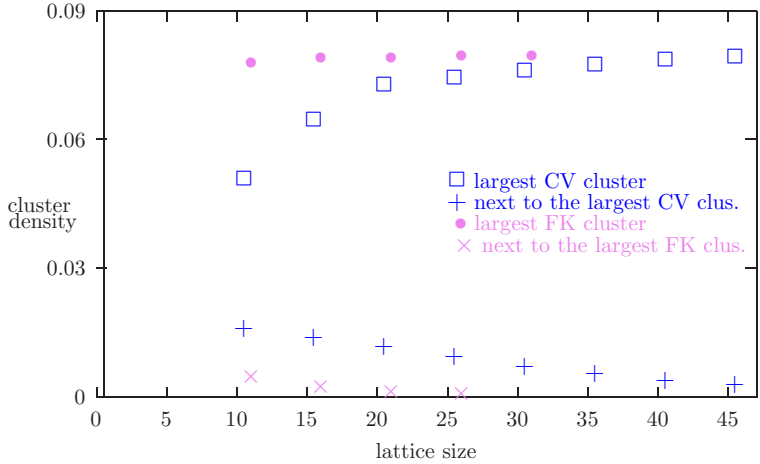} 
%\vspace{5cm}  % amount of vertical space needed
\caption{Density of the largest  center vortex (squares) and 
FK (circles) clusters. The densities of the second largest clusters
of the two types are also shown.}
\end{figure}
Thus the Wilson loop obeys a perimeter law in this phase. Conversely 
we can argue that a decay of large Wilson loops with an area law
implies the presence of an infinite 
FK cluster. Quite remarkably,  this kind of reasoning, based only on
concepts well defined at the microscopic level,  leads to the very important,
unavoidable, conclusion  that confinement 
\underline{requires}  monopole condensation. This is just the reverse 
of the usual 't Hooft confinement criterion, which can be justified
using plausibility arguments based on
 the effective description in terms of (dual) Abrikosov vortices. 

\par
The above arguments lead us to conjecture that the infinite FK cluster
encodes the most relevant part of the infrared dynamics of the
confined phase. More precisely, starting from the  the asymptotic
form of the vacuum  expectation value of a large rectangular Wilson
loop
\eq
\bra W(R,T)\ket =\exp\left(-\sigma RT+p(R+T)+c\right)
\left[\frac{\sqrt{R}}{\eta(\tau)}\right]^\um ~~~,~\tau=i\frac TR~,
\label{W}
\en   
where $\eta(\tau)$ is the Dedekind function and the term within square 
brackets is the universal contribution due to the quantum fluctuations
of the flux tube, it is easy to prove a sort of
no-renormalization theorem:
 the vacuum expectation value of $ W(R,T)$ in the
ensemble of configurations obtained  by those in thermal equilibrium
erasing all the finite FK clusters has exactly the same functional form of 
Eq.\ref{W} and with the same value of $\sigma$. The argument goes as
follows: let $N$ be the total number of configurations and $N_1$ those
compatible with $C$. We have $\bra W(C)\ket=\frac{N_1}N$. When the
finite FK clusters are eliminated, the number $N_1'$ of compatible 
configurations increases $(N_1'>N_1)$ because some graph with
$\varpi_C(G)=0$ is promoted to a compatible one. The perimeter law
decay produced by the finite FK clusters can now be used to get
$N_1=e^{-\alpha\, p(C)}N_1'$, which completes the proof.

The transformation $N_1\to N_1'$ reduces drastically the noise in the
evaluating of large Wilson loops and it has been used also in related
models to study the universal shape effects related to the quantum
fluctuations of the flux tube \cite{Gliozzi:2001tu}. 

\section{Clusters of center vortices}
Center vortices in $Z_2$ gauge model are constructed by assigning a 
vortex line in the dual lattice to each frustrated plaquette 
$(i.e.~ U_\Box=-1)$ in the direct lattice. Since the 
product of the six plaquettes forming a cube
is constrained to be equal to 1, the resulting graph of center
vortices (CV) in $\widetilde\Lambda$ has even coordination number. 
This has to be contrasted with the FK clusters of the previous 
section which are {\sl arbitrary} subgraphs of  $\widetilde\Lambda$.
\par
The value of a Wilson loop $W(C)$ in
a given configuration is $\pm 1$ according to the number, modulo 2, of
frustrated plaquettes of an arbitrary surface $\Sigma$ bounded by the 
loop ($\partial\Sigma=C$), or, 
in the center vortex language, to the number, modulo 2, of vortex 
lines that are linked to the loop. 
The role of center vortices in confinement in this
model is in a way trivial, since removal of  all center vortices
from $\Z_2$ gauge theory configurations simply removes all the
dynamics by transforming every configuration into the trivial
vacuum. Therefore center vortices are, in this sense, trivially
responsible not only for confinement, but for the whole dynamics of 
the model.
\par
We are interested to a subtler issue, namely the effect of the
finite CV  clusters  on confinement \cite{Engelhardt:1999fd}. 
Also in this case it is easy to prove that confinement requires an
infinite CV vortex, however it is no longer possible to prove the
no-renormalization theorem of the FK formalism, hence there is no
reason to believe that the string tension does not change under the
transformation erasing the finite CV clusters.

\par
It is straightforward to verify numerically that in the confining phase 
of $\Z_2$ gauge system there is no ambiguity in finding  a cluster of 
center vortices whose size scales linearly with the lattice volume for 
large enough lattices (see Fig.1), while in the deconfined phase we
found  that the density of the largest cluster decreases rapidly with
the volume.
   
It has to be noted that the presence of the infinite cluster does not 
necessarily imply a percolation property of the central
vortices. Confinement requires merely the presence of an infinite
cluster. In the FK representation this requirement is sufficient to
assure percolation of the infinite FK cluster, while there are
regions in the confining phase where the infinite CV  cluster does
not percolate, perhaps because of the even coordination constraint.

\begin{figure}
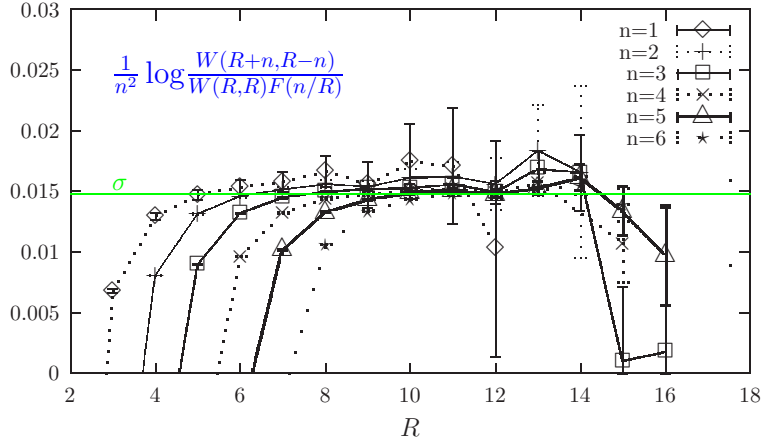
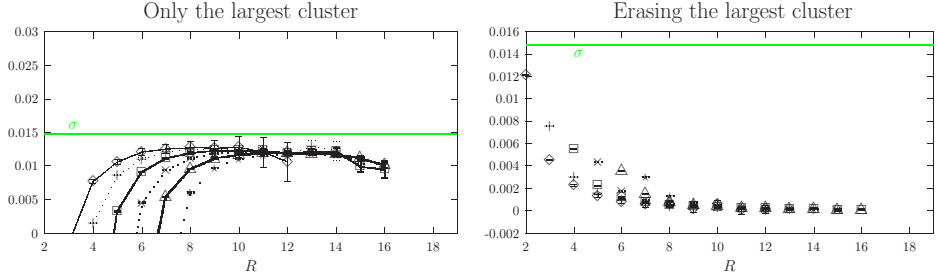

\label{f2}
\centering
\includegraphics[width=.8\textwidth]{fig2a.epsi}\\
\centerline{(a)}
\vskip.2 cm
\begin{minipage}[t]{.48\textwidth}
\centering
~
\includegraphics[width=\textwidth]{fig2b.epsi}\\
\centerline{ (b)}
\end{minipage}
%\vspace*{0.9cm}
\begin{minipage}[t]{.02\textwidth}
\centering
 ~
\end{minipage}
\begin{minipage}[t]{.48\textwidth}
\centering
\includegraphics[width=\textwidth]{fig2c.epsi}\\
\centerline{ (c)}
\end{minipage}
%
%\vspace*{0.3cm}
%
%\includegraphics[width=.5\textwidth]{fig2c.epsi}\\
%\centerline{ (c)}
%
%\end{minipage}
\caption{ (a)  $\sigma_{\rm eff}(R,n)$ as a
function of $R$ for $n=2,3,4,5,6$ for the original configurations. 
(b) Th same quantity  for the configurations where only the largest
cluster of center vortices has been left. (c) The effect of removing  
only the largest cluster.}
\end{figure}

 To study more carefully the relationship between the presence of the 
infinite cluster and the value string tension,
we chose  \cite{Gliozzi:2002ht}
to simulate the model at $\beta=0.74883$, which is well
inside the scaling region, and for which the value of the string
tension is known with high precision from simulations of the dual
model, that is the 3D (spin) Ising model 
\cite{Hasenbusch:1992zz,Caselle:1994df} $\sigma=0.01473(10)$
\par
First, we verified the {\em existence}
of an ``infinite'' cluster of vortex lines: for each configuration, we
selected the largest connected component of the graph defined by the
center vortices, and verified that the size of such component grows
linearly with the volume of the lattice.
\par 
The results of this analysis are shown in Fig. 1, where the size of
the largest CV cluster divided by the lattice volume is shown to approach
a constant for large lattices. The size of the second largest cluster
is also shown: its size relative to the lattice volume tends to
zero and the identification of the ``infinite'' cluster is
unambiguous. For comparison, also the corresponding FK cluster
densities are reported.

To test the relevance of the largest cluster of center vortices,
we proceeded as follows:
first, we modified each configuration in the Monte Carlo ensemble 
by eliminating  all the vortices not belonging to the largest
cluster, and second, by eliminating, instead, the largest cluster
only. The qualitative picture described above suggests that a non-zero
string tension will be found in the first case but not in the second.
\par

An efficient method to extract the string tension from Wilson loop
data generated by Monte Carlo simulations, which takes into account
the string fluctuation contribution, was introduced in
Ref.~\cite{Caselle:1996ii}. One
defines the ratio of the expectation values of rectangular Wilson
loops with the same perimeter:
\eq 
r(R,n)=\frac{\langle W(R+n,R-n)\rangle}{\langle W(R,R)\rangle}
\en
Using the asymptotic expansion defined in Eq.\ref{W} we get
\eq
r(R,n)\sim \exp\left({\sigma n^2}\right)\  F\left(n/R\right)~~,
\en
where $ F(t)=\sqrt{\eta(i)\sqrt{1-t}/\eta\left(
i\frac{1+t}{1-t}\right)}$,
so that one can define the following quantity
\eq 
\sigma_{\rm eff}(R,n)=\frac1{n^2}{\log\left(\frac{r(R,n)}{F(n/R)}
\right)}
\label{sigma}
\en
which approaches the string tension for large enough $R$.
The results of our simulations are reported in Fig.2
\par
In conclusion, our results confirm the picture of confinement as
due to the existence of an infinite cluster of center vortices: our
choice of the $\Z_2$ gauge theory allows us to bypass all the problems
related to the gauge-fixing and center projection that one encounters
when studying the same issue in $SU(N)$ gauge theories.
Two important new facts emerge from our study:
\par
While the largest center vortex is responsible for confinement,
since its removal from the configurations makes the string tension
vanish, the string tension measured from configurations in which all
the other clusters have been removed does not reproduce the full
string tension of the original theory. Therefore small clusters of
vortices, while unable by themselves to disorder the system enough to
produce confinement, do give a finite contribution to the string
tension of the full theory. This has to be contrasted with the FK 
cluster formulation, where the infinite cluster account for the whole 
string tension, because of a no-renormalization theorem.
\par
The quantum fluctuations of the flux tube survive the elimination of
the small CV or FK clusters: the Wilson loop after deletion of all the
small clusters show the same shape dependence as the ones of
the full theory, which can be explained as originating by the
fluctuations of a free bosonic string.


\begin{thebibliography}{99}
%\cite{'tHooft:1981ht}
\bibitem{'tHooft:1981ht}
't Hooft,G\ (1981) \ \
Topology Of The Gauge Condition 
And New Confinement Phases In Nonabelian Gauge Theories,
Nucl.\ Phys.\ B {\bf 190},  455.
%%CITATION = NUPHA,B190,455;%%

%\cite{tHooft:1977hy}
\bibitem{tHooft:1977hy}
't Hooft, G.\ (1978)\ \ 
        On the phase transition towards permanent quark confinement,
        Nucl.\ Phys.\ B {\bf B138}, 1
%%CITATION = NUPHA,B138,1;%%

%\cite{Giacomo:2002mm}
\bibitem{Giacomo:2002mm}
Di~Giacomo,A.,\ (2002)\ \
Color confinement and dual superconductivity: an update,
arXiv:hep-lat/0204032.
%%CITATION = HEP-LAT 0204032;%%

%\cite{Faber:2002ib}
\bibitem{Faber:2002ib}
Faber,M., Greensite,J. and Olejnik,S.,\ (2002) \ \
 Status of center dominance in various center gauges,
arXiv:hep-lat/0204020.
%%CITATION = HEP-LAT 0204020;%%


\bibitem{Gliozzi:2001tu}
Gliozzi, F. and Rago, A.,\ (2001)\ \
String breaking mechanisms induced by magnetic and electric  condensates,
Nucl.\ Phys.\ Proc.\ Suppl.\  {\bf 106}, 682
[arXiv:hep-lat/0110064].
%%CITATION = HEP-LAT 0110064;%%

%\cite{Engelhardt:1999fd}
\bibitem{Engelhardt:1999fd}
Engelhardt,M., Langfeld,K.,Reinhardt, H. and Tennert,O., \ (2000) \ \
Deconfinement in SU(2) Yang-Mills theory as a center vortex
percolation  transition,
Phys.\ Rev.\ D {\bf 61}, 054504
[arXiv:hep-lat/9904004].
%%CITATION = HEP-LAT 9904004;%%

%\cite{Hasenbusch:1992zz}
\bibitem{Hasenbusch:1992zz}
Hasenbusch, M., and Pinn, K.,\ (1992)
Surface tension, surface stiffness, and surface width of the
three-dimensional Ising model on a cubic lattice, 
Physica A {\bf 192}, 342
[arXiv:hep-lat/9209013].
%%CITATION = HEP-LAT 9209013;%%

%\cite{Caselle:1994df}
\bibitem{Caselle:1994df}
Caselle,M. Fiore,R.,Gliozzi,F., Hasenbusch,M.,Pinn, K. and Vinti,S. \ (1994)
Rough interfaces beyond the Gaussian approximation,
Nucl.\ Phys.\ B {\bf 432}, 590
[arXiv:hep-lat/9407002].
%%CITATION = HEP-LAT 9407002;%%

%\cite{Caselle:1999hy}
\bibitem{Caselle:1999hy}
Caselle, M. and Gliozzi, F. \ (2000)\ \
Thermal Operators in Ising Percolation,
J.\ Phys.\ A {\bf 33} , 2333
[arXiv:cond-mat/9905234].
%%CITATION = COND-MAT 9905234;%%

\bibitem{Gliozzi:2002ht}
Gliozzi,F.,Panero,M. and Provero,P. (2002)
Large center vortices and confinement in 3D Z(2) gauge theory,
arXiv:hep-lat/0204030.
%%CITATION = HEP-LAT 0204030;%%



%\cite{Caselle:1996ii}
\bibitem{Caselle:1996ii}
Caselle,M.,Fiore,R.,Gliozzi,F.,Hasenbusch,M. and Provero,P.(1997)
String effects in the Wilson loop: A high precision numerical test,
Nucl.\ Phys.\ B {\bf 486}, 245
[arXiv:hep-lat/9609041].
%%CITATION = HEP-LAT 9609041;%%



\end{thebibliography}
\end{document}